\documentclass[sigconf, review,anonymous=false]{acmart}

\AtBeginDocument{%
 \providecommand\BibTeX{{%
   \normalfont B\kern-0.5em{\scshape i\kern-0.25em b}\kern-0.8em\TeX}}}

\setcopyright{none}
\renewcommand\footnotetextcopyrightpermission[1]{}
\usepackage{amsmath,amsfonts}
\usepackage{algorithmic}
\usepackage{comment}
\usepackage{graphicx}
\usepackage{textcomp}
\usepackage{xcolor}
\usepackage{subfiles}
\usepackage{xurl}
\usepackage{array}
\usepackage{multirow}
\usepackage{xspace}
\usepackage{enumitem}
\usepackage{hyperref}
\usepackage{footmisc}
\usepackage{tabularx} 

\newcolumntype{L}{>{\centering\arraybackslash}m{1.5cm}}
\newcommand{\Sys}{Lockbox\xspace}
\newcommand{\App}{DOA App\xspace}
\newcommand{\design}{Zero Trust\space}

\begin{document}

\title{\Sys\ - A Zero Trust Architecture for Secure Processing of Sensitive Cloud Workloads}

% \author{Vamshi Krishna Thotempudi, Mahima Agarwal, Raghav Batta, Anjali Mangal}
% \email{{vthotempudi, mahagarwal, raghavbatta, anjalimangal}@microsoft.com}
% \affiliation{%
%  \institution{Microsoft Corporation}
%   \city{Mountain View}
%   \country{United States}
% }

\author{Vamshi Krishna Thotempudi}
\email{vthotempudi@microsoft.com}
\affiliation{%
 \institution{Microsoft Corporation}
  \city{Redmond}
  \country{United States}
}

\author{Mahima Agarwal}
\email{mahagarwal@microsoft.com}
\affiliation{%
 \institution{Microsoft Corporation}
  \city{Mountain View}
  \country{United States}
}

\author{Raghav Batta}
\email{raghavbatta@microsoft.com}
\affiliation{%
 \institution{Microsoft Corporation}
  \city{Mountain View}
  \country{United States}
}

\author{Anjali Mangal}
\email{anjalimangal@microsoft.com}
\affiliation{%
 \institution{Microsoft Corporation}
  \city{Mountain View}
  \country{United States}
}

\renewcommand{\shortauthors}{Thotempudi, et al.}
\begin{abstract}

Enterprises increasingly rely on cloud-based applications to process highly sensitive data artifacts. Although cloud adoption improves agility and scalability, it also introduces new security challenges such as expanded attack surfaces, a wider radius of attack from credential compromise, and challenges maintaining strict access controls across users, services, and workflows. These challenges are especially acute for applications that handle privileged data and execute security-critical analysis, where traditional trust boundaries and ad hoc safeguards are insufficient.

This paper presents \emph{\Sys}, a \design architecture designed for secure processing of sensitive cloud workloads under strict enterprise security and governance requirements. \Sys applies explicit trust verification, strong isolation, least-privilege access, and policy-driven enforcement throughout the entire application lifecycle, from user authentication and document ingestion to analysis execution and result storage. The system incorporates modern cloud security primitives including; role-based access control, centralized key management, encryption in transit and at rest, and controlled integration with cloud-based data processing services, ensuring that sensitive artifacts remain protected and accessible only to authorized users.

We discuss the usage of \Sys in processing highly sensitive cybersecurity reports and demonstrate how this architecture enables organizations to safely adopt advanced capabilities, including AI-assisted processing, without weakening their security posture.

\end{abstract}
% \begin{CCSXML}
% <ccs2012>
%  <concept>
%   <concept_id>00000000.0000000.0000000</concept_id>
%   <concept_desc>Do Not Use This Code, Generate the Correct Terms for Your Paper</concept_desc>
%   <concept_significance>500</concept_significance>
%  </concept>
%  <concept>
%   <concept_id>00000000.00000000.00000000</concept_id>
%   <concept_desc>Do Not Use This Code, Generate the Correct Terms for Your Paper</concept_desc>
%   <concept_significance>300</concept_significance>
%  </concept>
%  <concept>
%   <concept_id>00000000.00000000.00000000</concept_id>
%   <concept_desc>Do Not Use This Code, Generate the Correct Terms for Your Paper</concept_desc>
%   <concept_significance>100</concept_significance>
%  </concept>
%  <concept>
%   <concept_id>00000000.00000000.00000000</concept_id>
%   <concept_desc>Do Not Use This Code, Generate the Correct Terms for Your Paper</concept_desc>
%   <concept_significance>100</concept_significance>
%  </concept>
% </ccs2012>
% \end{CCSXML}

% \ccsdesc[500]{Do Not Use This Code~Generate the Correct Terms for Your Paper}
% \ccsdesc[300]{Do Not Use This Code~Generate the Correct Terms for Your Paper}
% \ccsdesc{Do Not Use This Code~Generate the Correct Terms for Your Paper}
% \ccsdesc[100]{Do Not Use This Code~Generate the Correct Terms for Your Paper}

%%
%% Keywords. The author(s) should pick words that accurately describe
%% the work being presented. Separate the keywords with commas.
\maketitle
%%%% Start remove header %%%%
\pagestyle{plain}
%%%% End remove header %%%% 

\section{Introduction}\label{sec:intro}
 % (\note{Vamshi to update the section label and fix the errors}) 

The recent advances in cloud computing have shifted enterprise workloads from a bounded perimeter to a highly distributed ecosystem of identities, APIs, managed services, blob storages, and third-party integrations. While this has significantly helped in agile development, segregation of concerns and improved the overall operational efficiency and uptime of such workloads, the traditional trust boundaries within internal networks like VPNs and static firewall zones have broken down and do not implicitly signal legitimacy. These trends expand the attack surface and increase the blast radius of a single compromised credential, misconfigured storage bucket, or over-privileged service account. The Red Hat State of Kubernetes security report~\cite{k8s-adoption} highlights that 67\% of companies are forced to delay or slow down Kubernetes application deployment in cloud environments due to security issues. The Flexera 2025 State of the Cloud Report~\cite{flexera-report} highlights that security remains a top challenge for 77\% organizations, second only to managing spend for migration of application workloads to cloud environment. In addition to security challenges impacting migration to cloud workloads, the applications already deployed in cloud environments are also experiencing significant increase in security challenges. According to the IDC Workload Cloud Migration and Placement Survey~\cite{idc-insights}, 89\% of organizations reported a year-over-year increase in cloud security incidents.  

\design architecture~\cite{zero-trust} responds to this reality by treating every access request as potentially hostile, continuously validating identity and device or workload posture, and enforcing least-privilege access, context-aware authorization for each transaction, instead of the whole traffic for a ``trusted`` network edge. This requirement becomes especially acute in industries like healthcare, finance and cybersecurity where cloud-hosted data is either high value or highly regulated. For example, in cybersecurity a compromise of internal threat intelligence like vulnerability assessments, or highly sensitive red team operation reports for advancing defenses can amplify ransomware impact and have broader impact across different organizations in the industry using the same cloud infrastructure and software packages. Industry research indicates that organizations recognize both the urgency and the difficulty of modernizing security for cloud-scale operations. The 2024 Gartner Information Technology survey~\cite{gartner-report} reported that 63\% of organizations worldwide have “fully or partially implemented” a \design strategy, reflecting broad adoption pressure as environments become more distributed and identity centric.  

This paper motivates \design for cloud environments by connecting these risk drivers to concrete architectural principles (continuous verification, least privilege, and explicit policy enforcement) and by showing how a \design approach can reduce the probability and consequences of data exposure.

The key contribution of this paper is to demonstrate a successful systematization of knowledge in building a highly secured system architecture for processing highly sensitive documents, adhering to the principles of \design leveraging latest advancements in secured cloud deployment using dual-key encryption, strong isolation and role based access control. It uses cloud Key Management Service ~\cite{kms-definition} as an enforcement mechanism for \design data access and bridges the gap between practical enterprise cloud operation and strong cryptographic confidentiality. This approach leverages what cloud platforms already do well (key isolation, audit logging) to cover what prior solutions left exposed (plaintext during processing), thereby markedly reducing
the attack surface in cloud workflows and addressing limitations identified in prior work.

% Next section layout
The rest of this paper is organized as follows. Section \ref{sec:related_work} reviews existing approaches and discusses their limitations. Section \ref{sec:design} describes the design principles that guided the development of \Sys, and the architectural decisions required to deploy such a system in a regulated enterprise environment. Section \ref{sec:case-study} demonstrates the usage of \Sys in processing of highly-sensitive Red Team reports and conducting AI-assisted processing while maintaining strict security safeguards. Finally, we conclude our paper and share key takeaways in Section \ref{sec:conclusion_future_work}.
\section{Related Work}
\label{sec:related_work}

Several works have explored client-side and hybrid encryption models for cloud data protection. Ahialey et. al.~\cite{Ahialey2025HybridEM}, demonstrate a hybrid RSA-AES encryption scheme for secure file storage on unreliable clouds. Chowdhury et. al.~\cite{10.1145/3548606.3560610}, present a dual-key approach to protect user content by encrypting data locally and storing an encrypted key in the cloud. While these methods effectively keep cloud operators from reading data at rest, they implicitly trust the cloud application to decrypt content for processing. \Sys reduces implicit trust by ensuring that cloud applications can decrypt content only within explicitly authorized contexts, triggered through a vault-mediated process. Unlike other encryption systems that assume servers will access plaintext once authorized, \Sys requires a key management service's approval for decryption, providing stronger authentication and authorization than previous client-side encryption models.

Recent research has also applied \design and cryptographic enforcement to development and operational workflows. Zmuda et.al.~\cite{11226761}, examine CI/CD pipeline security, noting that secrets and keys are potential leak points and recommending policy-driven controls, but their focus is on process and policy, not data encryption. Cheerla et. al.~\cite{11226461}, find that while fully homomorphic encryption and garbled circuits can keep data encrypted throughout ML inference, these methods are slow and complex, making them impractical for many time-sensitive cloud AI tasks. \Sys strikes a practical balance by limiting plaintext exposure and decrypting data only during analysis in a secure, audited environment. Unlike complex homomorphic encryption or secure enclaves, it leverages existing cloud tools like client-side encryption, key management service, and in-memory processing to deliver robust security without custom cryptography or hardware.

In summary, the existing approaches only address part of the problem space. \Sys brings these elements together, offering an integrated approach that unifies their strengths and provides a concrete solution with an architecture. The system however does not aim to eliminate all plaintext exposure, but to minimize and tightly control it using deployable cloud primitives.

\section{\Sys Methodology}
\label{sec:design}

\subsection{Design Goals}
\label{subsec:design-goals}
\Sys introduces a security first architecture for ingesting, processing, and analyzing sensitive documents within cloud based data processing systems. The methodology is framed by enterprise grade threat modeling assumptions and emphasizes rigorous controls across the entire data life-cycle. The approach centers on four foundational security goals.
% \subsubsection{Zero Trust Enforcement}

\begin{enumerate}
    \item {\bf \design Enforcement: }\Sys follows a strict \design posture, meaning every action - such as signing in, uploading information, or accessing data - requires explicit verification. No part of this design is automatically trusted.
    \item {\bf Dual Key Encryption Security: } \Sys employs a dual key protection model in which one key safeguards data on the user side while a service managed key provides an additional layer of control. Together, they ensure that information remains protected throughout transfer and storage, and that decryption occurs only within authorized environments.
    \item {\bf Strong Isolation for Sensitive Analysis: } \Sys ensures that sensitive information is handled only within a tightly controlled environment, plaintext documents never leave the application's controlled execution boundary.
    \item {\bf Secure Cloud Integration: } Any interaction with cloud based processing happens only after proper authorization, and all stored documents - both intermediate and final - are kept protected under strict access and lifecycle controls.
\end{enumerate}

\subsection{Application Architecture Overview}
\Sys follows a layered architectural model to correspond to the stages of secure document processing. Each layer is responsible for specific functionality as shown in Figure~\ref{fig:application_architrecture_layers} and is designed with security controls described in section ~\ref{subsec:design-goals}  

Below we provide a brief overview of this layered model for the secured deployment of the \Sys system: 
\begin{figure}[t]
    \centering    \includegraphics[width=0.48\textwidth]{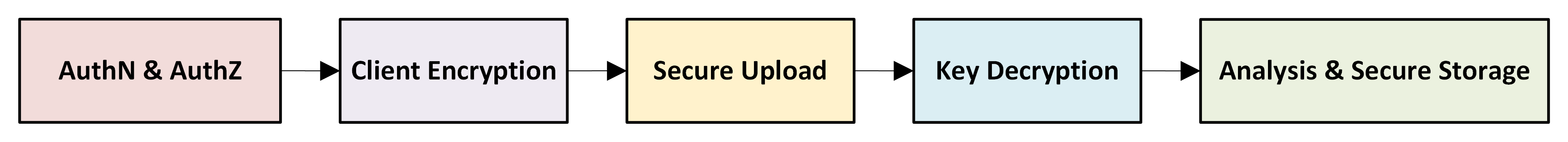}
    \caption{\Sys's Layered Model}
    \label{fig:application_architrecture_layers}
\end{figure}
\begin{enumerate}
    \item \textbf{User Access and Authorization Layer}: Users sign in and have their access verified before the system accepts any data, ensuring every action is tied to a validated identity.
    \item \textbf{Client Side Cryptography and Upload Layer}: Server generates an RSA key pair and shares the public key with the client. The document is secured on the user’s device before it leaves the browser. Only encrypted content is uploaded, so no readable data ever leaves the user’s machine.
    \item \textbf{Server Side Decryption and Data Processing Layer}: Server temporarily decrypts the data in memory only for processing. The readable content is never stored or exposed outside this tightly controlled area.
    \item \textbf{Secure Storage, Retention, and Result Access Layer}: Both the encrypted input and processed results are stored using access-controlled storage. Retention policies ensure that only authorized users can access the data and that sensitive information is purged after the designated period.    
\end{enumerate}
% \subsubsection{\bf User Access and Authorization Layer : }
% Users sign in and have their access verified before the system accepts any data, ensuring every action is tied to a validated identity. 

% \subsubsection{\bf Client Side Cryptography and Upload Layer : } Server generates an RSA key pair and shares the public key with the The document is secured on the user’s device before it leaves the browser. Only encrypted content is uploaded, so no readable data ever leaves the user’s machine.

% \subsubsection{\bf Server Side Decryption and Data Processing Layer : } A secure backend environment temporarily decrypts the data in memory only for processing. The readable content is never stored or exposed outside this tightly controlled area.

% \subsubsection{\bf Secure Storage, Retention, and Result Access Layer : } Both the encrypted input and processed results are stored using protected, access controlled storage. Retention policies ensure that only authorized users can access the data and that sensitive information is removed after the designated period.
\Sys's {\bf trust boundary} ensures that plaintext is exposed only within two tightly controlled zones: the user’s browser and the server’s ephemeral in‑memory analysis environment. All data flows across this boundary are cryptographically protected - client to server communication happens only through an encrypted route. 

\begin{figure*}[t!]
    \centering
    \includegraphics[width=\linewidth]{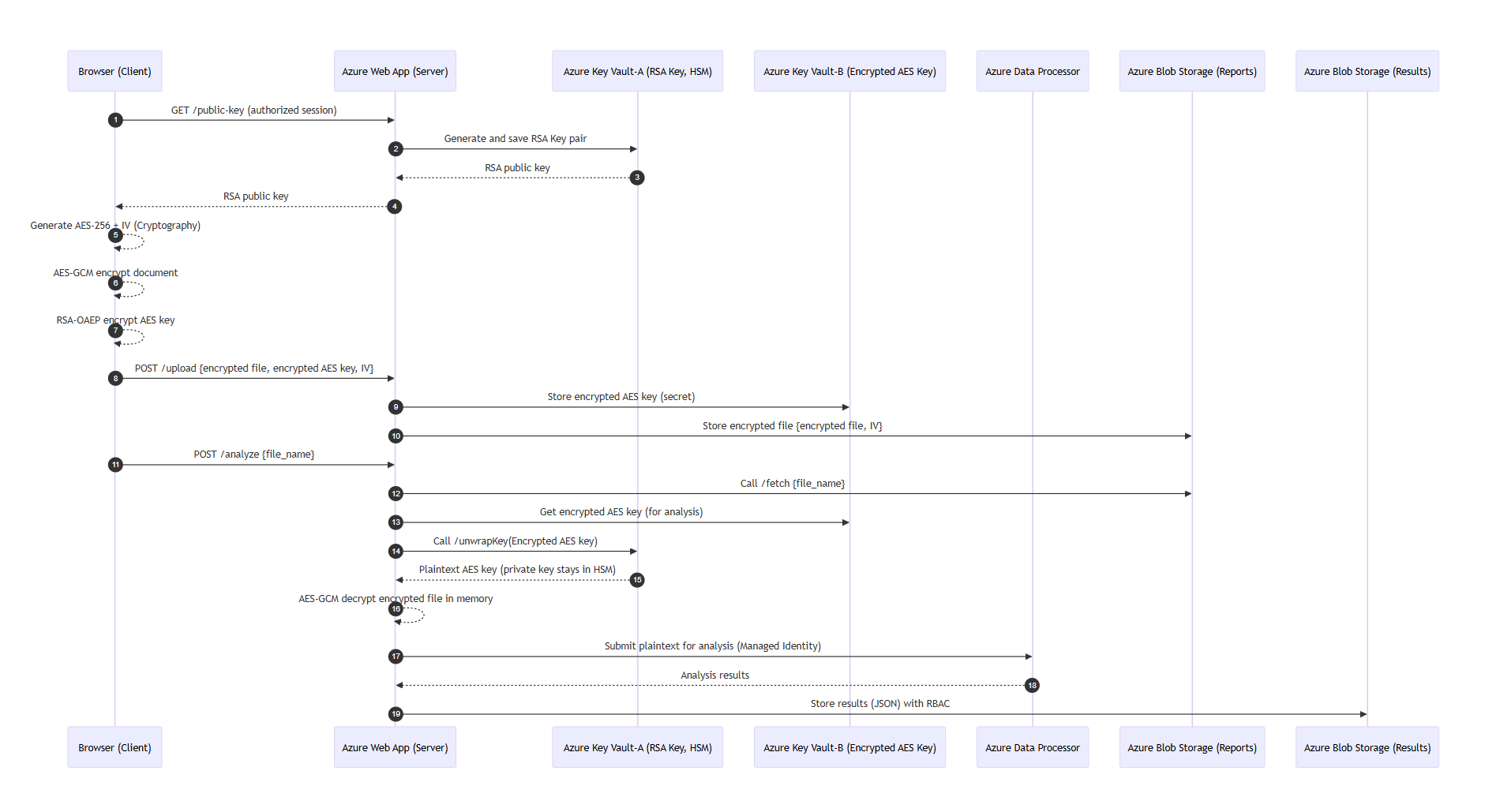}
    \caption{\Sys's Detailed Architecture}
    \label{lockbox_overview}
\end{figure*}
%%%%%%%%%%%%%%%%%%%%%%%%%%%%%%%%%%%%%%%%%
% 1. Detailed Architecture
%%%%%%%%%%%%%%%%%%%%%%%%%%%%%%%%%%%%%%%%%

\subsection{Detailed Architecture}
\label{detailedArchitecture}

In this section we provide a comprehensive breakdown of \Sys's architecture, demonstrating the interactions between different components, cryptographic operations and how data moves among the trust boundaries ensuring the design principles we already established in ~\ref{subsec:design-goals}. Figure ~\ref{fig:application_architrecture_layers} already outlines the layers in this architecture, and as we dive deeper, this can be broken down into sequence of steps, from user login to final data storage. 

Although the architecture of \Sys is designed to remain cloud-agnostic, we present it using Microsoft Azure as the reference platform, reflecting where the prototype was built and evaluated. The same model can be realized on any cloud provider that offers comparable isolation, identity, and secure storage capabilities.

\subsubsection{\bf Authentication and Authorization: } When the user tries to access \Sys's interface using a browser, if not already signed in, they are redirected to Azure Entra ID ~\cite{azure-entra} for authentication. Upon return, \Sys verifies the identity token and checks the user’s roles/permissions (only specific roles can upload or view results). Only if authentication and authorization succeed, user has access to upload functionality; otherwise, access is denied. \textit{This ensures that the encryption workflow begins in a trusted context.} 

\subsubsection{\bf RSA Key Pair Generation (via Key Vault): } When an authorized user initiates an upload, server requests an instance of Azure Key Vault service~\cite{key-vault} to create an RSA key pair. The key pair is stored as a Key Vault ``Key'' object (non-exportable), meaning the private key will remain inside the key vault (\textit{Key Vault-A}) and cannot be directly retrieved. Key Vault-A returns the public key (safe to share) to the server and then sends it to the user’s browser (in PEM format) for use in client-side encryption. At this point, this design architecture has:
\begin{itemize}
    \item An RSA public key on the client, tied to a private key which remains secure in Key Vault-A.
    \item Confidence that only our managed service identity can use the private key (via Key Vault’s role-based policies).
\end{itemize}

\subsubsection{\bf Client-Side Encryption \& Upload: } \label{subsubsec:client-side-enc-upload} Using the received public key, the user’s browser now performs encryption of the document. The client (browser) generates a random 256-bit AES key with a random IV (12 bytes) which is used to encrypt the document. Next, the browser RSA-OAEP encrypts the AES key using the RSA public key, producing an encrypted AES key. The browser then uploads:
\begin{itemize}
    \item AES-encrypted document
    \item RSA-encrypted AES key and IV
\end{itemize}

Above data is sent via an HTTPS POST to the \Sys server and server uploads the encrypted document and IV as metadata to Azure Blob Storage~\cite{blob-storage} which is encrypted and secured with RBAC policy. RSA-encrypted AES key gets saved into second key vault (\textit{Key Vault-B}).
At this stage, no plaintext document or unencrypted key has left the user’s device. Even if intercepted, the data is useless without the private key. 

\subsubsection{\bf Key Unwrapping and Document Decryption (Server-Side): } When the user initiates data processing, the server immediately proceeds to decrypt the file in a secure manner:
\begin{itemize}
    \item Server calls Azure Key Vault-A’s unwrapKey API ~\cite{unwrap-key} with the encryption key (the RSA wrapped AES key) from Key Vault-B and references the appropriate RSA private key. Key Vault uses the RSA private key internally, without exposing it to the application, as documented by the Key Vault cryptographic service design \cite{azure-key-hsm}, to decrypt encryption key. It returns the plaintext AES key to the server. Server is granted the decrypted AES key but never handles the raw RSA private key, eliminating the risk of private-key exposure in application memory.
    \item With the 256-bit AES key now available (and the IV from the file metadata), the server decrypts the encrypted file entirely in memory using AES-GCM. The GCM authentication tag is verified, ensuring the ciphertext wasn’t tampered with. On success, the original plaintext document is reconstructed in the server’s memory. The server keeps this plaintext only transiently, for immediate analysis.
    \item \textit{Trust boundary:} The sensitive AES key now existed in the server’s memory for a brief duration during decryption and the design ensures this happens only after user authentication; inside a controlled trusted boundary.
\end{itemize} 

\subsubsection{\bf Analysis of Decrypted Content: } As soon as the document is decrypted, the plaintext content is handed off to \textbf{Azure Data Processor}, authenticated using the Managed Identity of the Azure Web App \cite{web-app} for the authentication. The plaintext is transmitted securely (TLS) to the Azure Data Processor, which processes data in-memory and returns only the analysis results to \Sys. The server receives these results (example, a JSON with findings or a generated summary). At this point, the plaintext document is no longer needed in memory. \Sys frees or scrubs those buffers as soon as analysis completes, minimizing any residual presence of sensitive data on the server. 

\subsubsection{\bf Secure Storage of Outputs: }After analysis, \Sys retains evidence and results in a secure form:
\begin{itemize}
    \item The encrypted document will be stored in Azure Blob Storage for a limited time. This allows optional re-analysis or auditing if needed, without exposing the content.
    \item The analysis results are stored in a separate Blob storage and are typically less sensitive than the full report (they might be abstracted insights), but they can still contain sensitive elements. Therefore, this blob storage is also protected by Azure RBAC policy and the data is encrypted at rest by default. Only users with authorization can read the results.
    \item Both blob storages have data retention policies. For example, the encrypted files can be automatically purged after 7 days, and results after 90 days, to reduce long-term risk.
\end{itemize}
\subsubsection{\bf Cleanup and Monitoring: } 
After decryption and processing complete, \Sys immediately clears all sensitive material from memory. The unwrapped AES‑256 key and decrypted document exist only briefly in server's RAM; no plaintext or key material is written to disk, logged, or cached. All buffers used during decryption are released, minimizing the exposure window.
If each document had a unique RSA key pair, those keys can later be disabled or deleted from Key Vault as per retention policy (though being non-exportable, they pose little risk if left). Audit logs in Azure Entra ID, Key Vault, and Blob Storage each contribute to the trail of who initiated an analysis, including each unwrapKey operation details such as the timestamp and the identity that performed the operation. These can be reviewed for any anomaly (like an unexpected unwrapKey call, which could indicate misuse). The robust logging and monitoring complements the preventive controls by offering detailed insights.
\section{Case Study: Detection Opportunity Assessment App Deployment}\label{sec:case-study}
In this section, we demonstrate the application of \Sys's \design architecture in a high-risk security use case, with the deployment of a Detection Opportunity Assessment application. We refer to this application as the \emph{\App}. The goal of this application is to processes highly sensitive red team reports, identify detection gaps and accelerate defensive response. We illustrate how \Sys enforces explicit trust verification, strong isolation, least-privilege access, and policy-driven security controls across the full lifecycle of sensitive artifacts.

\subsection{Use Case Background}
As part of ongoing efforts to strengthen cybersecurity posture and reduce attack surface, organizations conduct structured adversary simulation exercises involving \textit{offensive (Red Team)} and \textit{defensive (Blue Team)} security teams. The red team specializes in simulating real-world attacks by identifying vulnerabilities, exploiting these through techniques such as phishing, credential theft, privilege escalation, and lateral movement, and documenting their activities in detailed reports. These reports describe attack execution paths, techniques used, vulnerabilities exploited, and gaps in existing defenses. As these reports provide explicit instructions for breaching enterprise infrastructure, they are classified as highly confidential. Unauthorized disclosure could enable adversaries to reproduce the same attacks with minimal effort.
The blue team consumes these reports to identify detection opportunities, author new detections, and deploy them to Security Operations Centers (SOC) to prevent similar attacks in the future. However, manual analysis of red team reports is time-consuming and labor-intensive. Depending on document complexity, the process can take several weeks from report delivery to detection deployment. During this interval, the organization remains vulnerable to these attacks if a real adversary independently discovers and exploits the same vulnerabilities.
To reduce this window of exposure, the \App, which uses AI-based analysis to rapidly extract adversary techniques, map them to established frameworks such as MITRE ATT\&CK \cite{mitre-matrix}, identify coverage gaps in existing detections, and generate analysis outputs for blue team review is deployed. While the application incorporates significant technical capabilities, this paper focuses on how \Sys's \design principles were applied to ensure its secure deployment. The functional and algorithmic details of the application ~\cite{doa-app} are beyond the scope of this work.

\subsection{Threat Model and Security Requirements}
The deployment of the \App introduces significant security challenges. Red team reports contain explicit attack paths, tooling details, and vulnerability information that must not be accessible to unauthorized users or services. Traditional application security models that assume trust within organizational boundaries are insufficient in this context. Any compromise of user credentials, application components, or supporting infrastructure could lead to unauthorized access to reports, resulting in severe security consequences. Following are the application's security requirements:
\begin{itemize}
\item \textbf{No implicit trust}: All access to red team reports requires explicit authentication and authorization.
\item \textbf{Strong isolation}: Red team reports must be isolated during upload, storage, processing, and retrieval.
\item \textbf{Controlled decryption}: Documents remain encrypted by default and are decrypted only within authorized analysis contexts.
\item \textbf{Auditability}: All access to reports, keys, and analysis results must be logged, with enforced retention policies.
\end{itemize}
\Sys was designed to satisfy these requirements by enforcing \design principles across the entire application lifecycle and does not address physical attacks on cloud infrastructure or side‑channel attacks against hardware security modules.

\subsection{User Roles and Workflows}
The \App supports two distinct user roles with different responsibilities and access privileges: red team users and blue team users. \Sys enforces strict role separation and ensures that each role can access only the resources necessary for their function.
\subsubsection{Red Team Users} Red team users are responsible for uploading newly completed adversary simulation reports to the application for analysis. The workflow proceeds as follows:
\begin{enumerate}
    \item \textbf{Authentication and Authorization}: When a red team user accesses the application, \Sys authenticates the user via the organization's identity provider and verifies that the user is authorized to upload reports. Authorization decisions are based on role membership and are enforced at the application entry point.
    \item \textbf{Document Protection}: Once authorized, the user selects a report from their local system. To ensure that the document is never transmitted or stored in plain text, the document is encrypted locally on the user's device using client-side tooling provided by the application before upload. The application provides client-side tooling to generate a per-document symmetric encryption key, encrypt the report content, and protect the symmetric key using a public key provided by \Sys. This ensures that even if network traffic is intercepted or storage is compromised, the plain text document remains inaccessible.
    \item \textbf{Encrypted Upload and Storage}: The encrypted document and the encrypted symmetric key are transmitted to the application backend. The encrypted document is stored in a dedicated, access-controlled storage with a retention policy that automatically deletes documents after a defined period (typically seven days). The encrypted symmetric key is stored separately in a secure key vault. \Sys uses managed identity to authenticate all interactions with storage and key management services, ensuring that credentials are never embedded in application code or configuration.
\end{enumerate}

\begin{figure}[t]
    \centering
    \includegraphics[width=1\linewidth]{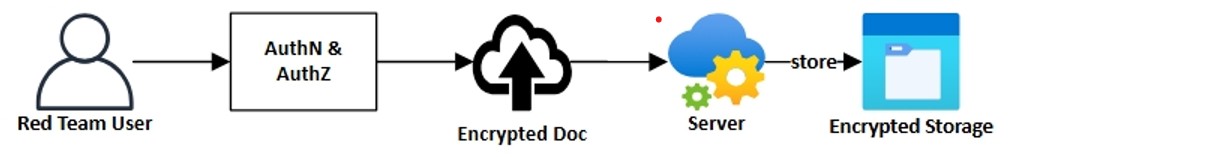}
    \caption{\App's Red Team User Workflow}
    \label{fig:red_team_workflow}
\end{figure}

Figure~\ref{fig:red_team_workflow} represents how red team users interact with \App.

\begin{figure}[t]
    \centering
    \includegraphics[width=1\linewidth]{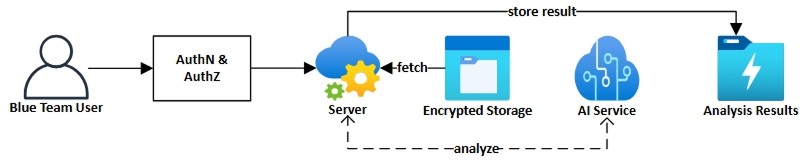}
    \caption{\App's Blue Team User Workflow}
    \label{fig:blue_team_workflow}
\end{figure}

\subsubsection{Blue Team Users} Blue team users are responsible for reviewing analysis results generated by the application and using them to author (or enhance) detections. The workflow proceeds as follows:

\begin{enumerate}
    \item \textbf{Authentication and Authorization}: Blue team users authenticate via the same identity provider as red team users. \Sys verifies that the user is authorized to specific reports based on role membership and organizational policy.
    \item \textbf{Analysis Execution}: Once authorized, the application displays a list of specific reports that the user has access to. For each accessible report, the user may initiate analysis. When analysis is initiated, the application backend retrieves the encrypted symmetric key from the key vault, decrypts it using a private key that never leaves the key vault (via the unwrapKey API), and uses the decrypted symmetric key to decrypt the report content. The decrypted document is then transmitted to an AI Service for analysis under strict security controls. Analysis results are stored in a separate, access-controlled storage and made available to authorized blue team users.
    \item \textbf{Access to Analysis Results}: The application displays analysis results only for reports on which the user has initiated analysis. Each result is associated with a specific red team report and includes extracted techniques, identified gaps, and recommended detections. Access to analysis results is logged for audit purposes.
\end{enumerate}

Figure~\ref{fig:blue_team_workflow} represents how blue team users interact with \App.

% Figure~\ref{fig:blue_team_workflow} is a representation of how the blue team user interacts with the \App.

\Sys ensures that decryption and analysis occur only after successful authentication and authorization, and that all cryptographic operations involving private keys are performed within the hardware-backed security boundary of the key vault.
\subsection{Document Protection and Processing}
The security architecture for document handling in the \App is designed to eliminate implicit trust and enforce defense in depth across the document lifecycle. This is achieved through a layered model combining client-side encryption, key isolation, controlled decryption, and policy-driven access control. 

\subsubsection{Client-Side Encryption}
As described in section ~\ref{subsubsec:client-side-enc-upload} documents are encrypted on the client side before upload using a per‑document symmetric key, ensuring that plaintext data never traverses the network.
\subsubsection{Key Management and Isolation}
\Sys uses Azure Key Vault to manage asymmetric key pairs and protect symmetric encryption keys. When the application is initialized, it requests the key vault to generate a non-exportable RSA key pair. The private key is stored as a cryptographic key object within the vault and is never exposed to the application or any other external entity. The public key is retrieved and provided to the client for encrypting symmetric keys.
When the application needs to decrypt a document for analysis, it retrieves the encrypted symmetric key from a separate key vault instance and sends it to the key vault holding the RSA private key. The vault's unwrapKey API performs the decryption operation internally and returns only the decrypted symmetric key to the application. This design ensures that the private key never leaves the hardware security module boundary, significantly reducing the risk of key compromise.
\subsubsection{Controlled Decryption and AI Analysis}
Decryption and analysis occur only within the application backend, and only after successful user authentication and authorization. Once the symmetric key is decrypted, the application uses it to decrypt the report content in memory. The decrypted document is then transmitted to a cloud-based AI analysis service using secure, authenticated channels. The AI service processes the document and returns structured analysis results, which are stored in a dedicated, access-controlled blob storage.
The application does not retain decrypted documents beyond the duration required for analysis, and all temporary artifacts are securely erased from memory. The AI service is configured to operate in a mode that does not retain customer data beyond the processing session, ensuring that sensitive documents are not inadvertently stored or logged by external services.
\subsubsection{Retention Policies and Lifecycle Management}
Both encrypted documents and analysis results are subject to automated lifecycle policies enforced by the storage infrastructure. Encrypted documents are automatically deleted after a configurable retention period (default: seven days), while analysis results are retained for a longer period to support blue team workflows. Access to both categories of artifacts is governed by role-based access control policies, and all access events are logged for audit and compliance purposes.

The deployment of the \App demonstrates how \Sys enables secure operation of high-risk security workloads that process extremely sensitive artifacts. By enforcing \design principles, strong isolation, least-privilege access, and end-to-end encryption, \Sys ensures that red team reports remain protected throughout their lifecycle, from client-side encryption and secure upload to controlled decryption, AI-assisted analysis, and policy-driven deletion. The architecture eliminates implicit trust, minimizes the blast radius of potential compromise, and provides the auditability and lifecycle controls required for regulated enterprise environments. This case study illustrates the practical application of the design principles discussed earlier in this paper and validates \Sys's ability to support secure deployment of advanced security capabilities in production settings, following internal threat modeling and security review.
% \documentclass[paper.tex]{subfiles}
% \begin{document}

\section{Conclusion}
\label{sec:conclusion_future_work}

\Sys demonstrates that a \design architecture, when combined with client-side AES-GCM encryption, non-exportable RSA key objects, and Key Vault-mediated unwrapKey operations, can securely process sensitive cloud workloads. By ensuring that plaintext exists only ephemerally in backend memory and only after identity verified key release, \Sys reduces the attack surface significantly compared to traditional cloud data processing pipelines, where plaintext is typically available throughout the service boundary. The architecture establishes a cryptographic gate: no other service can access sensitive content unless an authenticated unwrapKey event occurs, enabling traceable enforcement of \design policy. 

\textbf{Key Takeaways \& Future Work:}
Future work includes the adoption of HSM backed RSA keys~\cite{azure-key-hsm} in Azure Key Vault Premium, providing FIPS 140-3 Level 3 ~\cite{hsm-fips} guarantees, stronger resistance to physical tampering, and a hardware anchored trust root for unwrapKey operations. Additional hardening includes establishing a key rotation policy for rotating RSA key pairs, and retiring old versions based on enterprise data sensitivity policies. Together, these enhancements strengthen \Sys's cryptographic posture and extend its applicability to regulated, high assurance environments.

% \newpage

%% The next two lines define the bibliography style to be used, and
%% the bibliography file.
%%% -*-BibTeX-*-
%%% Do NOT edit. File created by BibTeX with style
%%% ACM-Reference-Format-Journals [18-Jan-2012].

% \bibliographystyle{ACM-Reference-Format}
% \bibliography{references}
%% If your work has an appendix, this is the place to put it.
\appendix

\section{Disclaimers}

\begin{enumerate}
    \item \textbf{Author Employment Notice}: All authors are employees of Microsoft Corporation.
    \item \textbf{Independence Notice}: The contents of this paper are provided for academic and research purposes.
    
    \item \textbf{Employee Disclaimer}: This paper represents the views of the authors and does not necessarily represent the views of Microsoft. This document does not describe a commercially available product and is for research and academic discussion only.
    \item \textbf{No-Promise / No-Warranty Disclaimer}: The techniques described are experimental and provided without warranty. Security claims depend on correct implementation and configuration.
    \item \textbf{Non-endorsement / Comparative Claim Disclaimer}: Mentions of third‑party products or technologies are for explanatory purposes only and do not constitute endorsement.
    \item \textbf{Data Disclaimer}: This paper does not contain or rely on any Microsoft proprietary data.
    \item \textbf{Microsoft Internal Technology Names}: This architecture uses Azure cloud services in accordance with publicly documented capabilities.
    \item \textbf{Export Controls / Cryptography Notice}: This document describes cryptographic methods. Export of cryptographic information is subject to applicable laws and regulations.
    
\end{enumerate}

% \note{
% [Raghav - Confirm the use of word marks ("Azure", "Key Vault", "Entra ID") follows Microsoft Brand guidelines. Update related work for citations also. Update Appendix
% Mahima - Cut out text from Section 3.2. Also increase font on Figure 1 and make it readable.
% Vamshi - Make the red and blue team diagrams clear and shorter]
% }

\end{document}